\documentclass[floatfix,aps,twocolumn,pra,superscriptaddress,notitlepage]{revtex4-1}
\usepackage{graphicx}
\usepackage{dcolumn}
\usepackage{bm}
\usepackage[T1]{fontenc}
\usepackage[utf8]{inputenc}
\usepackage{textcomp}
\usepackage{amsmath}
\usepackage{mathrsfs}
\usepackage{amssymb}
\usepackage{units}
\usepackage{xcolor}
\usepackage{float}
\usepackage{mathtools}
\usepackage{array,multirow,makecell}
\usepackage{tabularx,colortbl}
\usepackage{hhline}

\DeclarePairedDelimiterX\braket[2]{\langle}{\rangle}{#1 \delimsize\vert #2}
\usepackage{physics}
\newcommand\hidden[1]{} 
\newtagform{hidden}[\hidden]{}{}
\newcommand{\iac}[1]{{\color{black}#1}}

\begin{document}


\title{
Linearized theory of the fluctuation dynamics in 2D topological lasers}

\author{Aurelian Loirette--Pelous}
\affiliation{Université Paris-Saclay, Institut d'Optique Graduate School, CNRS, Laboratoire Charles Fabry, 91127, Palaiseau, France}
\affiliation{INO-CNR BEC Center and Dipartimento di Fisica, Università di Trento, 38123 Povo, Italy}
\author{Ivan Amelio}
\affiliation{INO-CNR BEC Center and Dipartimento di Fisica, Università di Trento, 38123 Povo, Italy}
\author{Matteo Seclì}%
\affiliation{International School for Advanced Studies (SISSA), Via Bonomea 265, I-34136 Trieste, Italy}
\author{Iacopo Carusotto}%
\affiliation{INO-CNR BEC Center and Dipartimento di Fisica, Università di Trento, 38123 Povo, Italy}

\date{\today}

\begin{abstract}
We theoretically study the collective excitation modes of a topological laser device operating in a single-mode steady-state with monochromatic emission. We consider a model device based on a two-dimensional photonic Harper-Hofstadter lattice including a broadband gain medium localized on the system edge. Different regimes are considered as a function of the value of the optical nonlinearity and of the gain relaxation time. The dispersion of the excitation modes is calculated via a full two-dimensional Bogoliubov approach and physically interpreted in terms of an effective one-dimensional theory. Depending on the system parameters, various possible \iac{physical processes leading to} dynamical instabilities are identified \iac{and characterized. On this basis, strategies to enforce a stable single-mode topological laser operation are finally pointed out}.
\end{abstract}

\maketitle


\section{\label{sec:level1}Introduction}
One of the most exciting applications of topological photonics are the so-called topological lasers~\cite{Harari2016,harari2018,Pilozzi2016,Solnyshkov2016}. Such {\em topolaser} devices are based on a topological photonic system embedding a suitable gain material, so that laser oscillation is induced to occur in a topologically protected edge mode~\cite{Ozawa2019,Ota2020}. 
So far, topolasing operation has been experimentally demonstrated both in the zero-dimensional edge states of one-dimensional arrays~\cite{stjean2017,Parto2018,Han2019,Ota2018} as well as in the one-dimensional edge modes of two-dimensional lattices~\cite{bahari2017,bandres2018,zeng2020electrically}. As it was theoretically pointed out~\cite{Wittek2017,harari2018}, such devices hold a promise for optoelectronic applications, since the chiral nature of the edge modes guarantees an efficient phase-locking of the emission over macroscopic distances as well as enhanced robustness against fabrication disorder~\cite{harari2018,amelio2019b}. This is of crucial importance whenever one needs to combine high power and long-lasting coherence in a single device.

While a clean single-mode emission has been achieved in~\cite{bahari2017,bandres2018}, several other experimental and theoretical works have pointed out more complex behaviours. The topological quantum cascade laser of \cite{zeng2020electrically} displays some secondary spectral peaks. For a tight-binding topolaser model,  the possibility of dynamical instabilities arising from the interplay of optical nonlinearities and slow carrier dynamics has been numerically highlighted \cite{longhi2018}. Since such effects may dramatically affect the coherence properties of the topolaser emission as well as its power efficiency, it is of crucial importance to fully understand the various \iac{processes that may lead to} instabilit\iac{ies}. 

In this work, we report a numerical and analytical study of the dispersion of the collective excitations around a monochromatically oscillating steady-state. Our study is based on the Bogoliubov theory of the collective excitations on top of dilute Bose-Einstein condensates~\cite{BECbook}, which was then generalized to lasers and non-equilibrium condensates of exciton-polaritons~\cite{wouters2007}. On one hand, our analysis allows identify\iac{ing} the general features of the excitation modes and the dynamics of quantum and classical fluctuations of generic topolaser devices. In particular, it provides microscopic support to the \iac{numerical observations in~\cite{secli2019} and to the} study of the long-distance and long-time correlators of the fluctuations that are involved in the spatio-temporal coherence properties of the emission~\cite{amelio2019b}. On the other hand, our theory \iac{recovers the dynamical instabilities anticipated in~\cite{longhi2018} and shines light on the different physical processes} that may \iac{destabilize a} monochromatic topolaser operation and, eventually, \iac{lead to} a chaotic multi-mode emission. \iac{A related study of the collective excitations of topolaser devices has appeared in~\cite{zapletal2020}, focusing on the case of a photonic Haldane model but restricting to the idealized class-A limit of a fast carrier dynamics.}

\iac{Here we go beyond this approximation and develop a more sophisticated theory that includes the slow carrier dynamics of realistic semiconductor-based devices.}  
While the idealized tight-binding model considered in \iac{the present work is likely to only provide qualitative insight on semiconductor laser arrays~\cite{pick2015}, we expect it to be quantitatively} predictive for the lattices of micropillars~\cite{baboux2018} used in polariton-based topolaser devices~\cite{klembt2018}. 
\iac{From a general theoretical perspective, our work offers a powerful framework of
\iac{major}
utility to characterize instability processes in generic topolaser systems. This} will be of great importance in view of designing devices where instabilities are tamed and the emission is robustly clean and monochromatic.

The structure of the work is the following. In Sec.~\ref{sec:model} we review the general concepts of a topological laser device based on including gain into a photonic topological Harper-Hofstadter model \iac{and we introduce the theoretical model. In Sec.~\ref{sec:2D_steady_state}, we characterize the steady-state of the lasing device.} In Sec.~\ref{sec:classA} we calculate the collective excitation modes in the \iac{simplest} regime where the gain medium has a very fast recovery time and no optical nonlinearity is present beyond gain saturation, \iac{finding a stable topolaser behaviour. An effective analytical 1D} theory able to recover the main features of the \iac{numerical} 2D calculation is \iac{then} proposed and quantitatively validated. In Sec.~\ref{sec:classB}, we \iac{extend our theory of collective excitations to more complicated regimes displaying a slow carrier dynamics in the gain medium and/or significant nonlinearities: this allows \iac{us} to identify the main processes that may lead to dynamical instabilities and to propose strategies to tame them.} 
Conclusions are finally drawn in Sec.~\ref{sec:conclu}.

\section{\label{sec:model} The Harper-Hofstadter topological laser}

In this Section, we review the general features of a laser device built by introducing gain on the edge of a topological photonic lattice. Going beyond our previous works~\cite{secli2019,amelio2019b}, we consider a wider class of devices where different regimes of operation are found depending on the timescale of the carrier dynamics in the gain medium and on the optical nonlinearity of the platform.

\begin{figure}[htbp]
    \centering
    \includegraphics[width=\columnwidth]{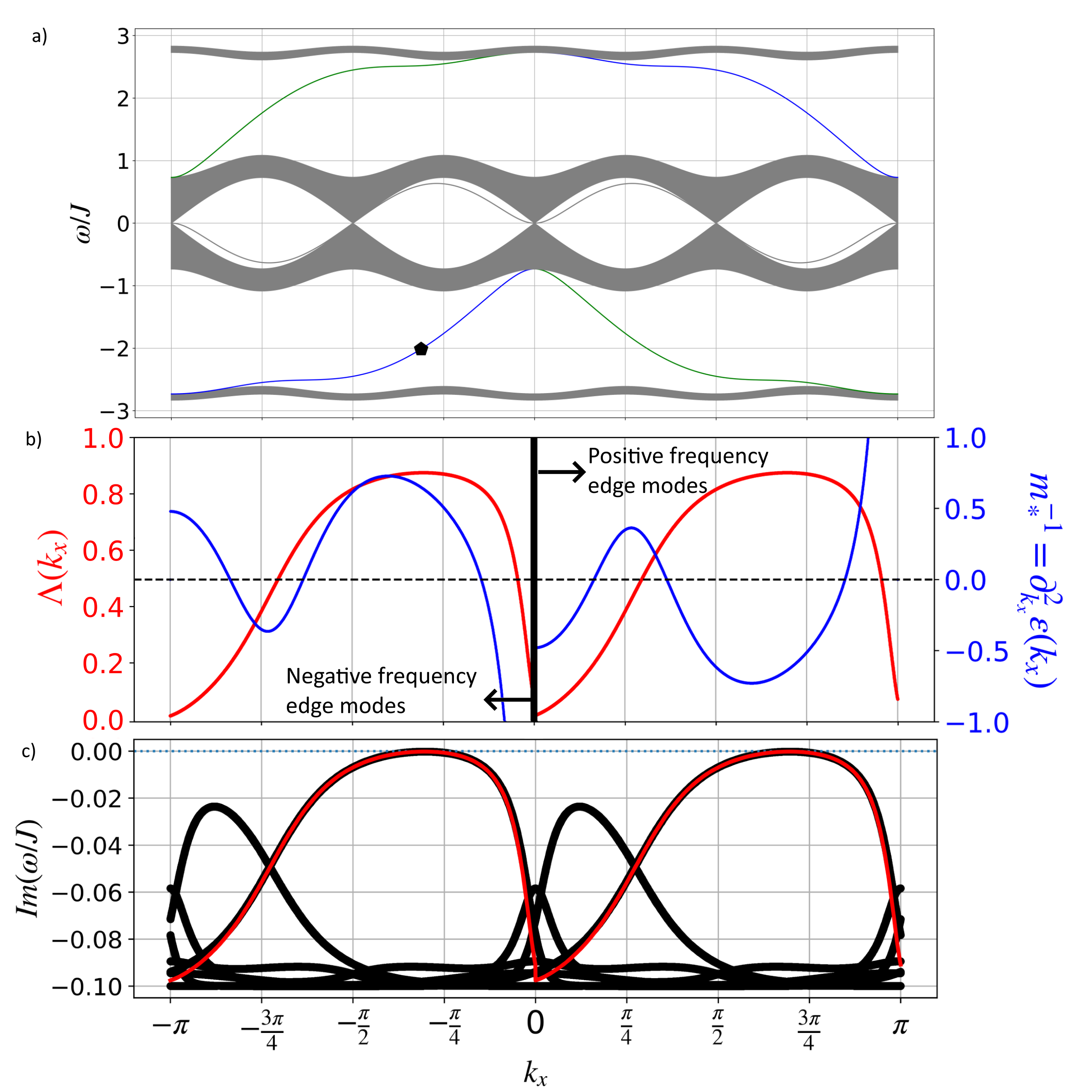}
    \caption{ {\it Harper-Hofstadter model.}  Panel (a): energy bands of the conservative Harper-Hofstadter Hamiltonian Eq.~(\ref{eq:HH}) with flux $\theta = 1/4$ in a finite lattice of $n_y = 399$ sites along $y$ with periodic boundary conditions along $x$. The blue (green) lines indicate the dispersion of the edge modes localized on the $y=1$ ($y=n_y$) edge. The dark dot indicates the spatially most localized edge mode within the lower energy gap on the $y=1$ edge.
    Panel (b): spatial localization function $\Lambda(k_x)$ (red line) and curvature of the dispersion (blue line) of the edge states. The left/right part of the plot refers to the edge mode living on the $y=1$ edge in the lower/upper energy gap. 
    Panel (c): imaginary part of the Bogoliubov spectrum of the linearized dynamics around the vacuum solution for a 
    pump strength right at the lasing threshold. 
    \iac{\iac{The t}hick black lines show the prediction of the full 2D model, while \iac{the} thin red ones show the prediction of the 1D effective theory for the excitations living on the $y=1$ edge.
    }
    }
    \label{fig1}
\end{figure}

\subsection{The Harper-Hofstadter model}
As a specific and most relevant example, we focus on the case of a photonic lattice implementing the so-called Harper-Hofstadter (HH) model~\cite{harper1955,hofstadter1976,Ozawa2019}. In the Landau gauge, the HH Hamiltonian  reads:
\begin{equation}\label{eq:HH}
    H =  - J \sum_{x,y} \Big\lbrace    \hat{\psi}^{\dagger}_{x,y} \hat{\psi}_{x,y+1} 
    + e^{-2\pi i \theta y} \hat{\psi}^{\dagger}_{x,y} \hat{\psi}_{x+1,y} + \textrm{h.c.} \Big\rbrace     
\end{equation}
where the zero of energies is set at the bare frequency of the sites, the sum runs over all the sites of the lattice, $x,y=1,...,n_{x,y}$,
$\hat{\psi}_{x,y}$ is the (bosonic) photon annihilation operator at the site $(x,y)$ and $J$ is a real-valued hopping amplitude.  The topological properties of this lattice are due to the synthetic magnetic field piercing it. Its strength is quantified by the flux $\theta$ per plaquette in units of the magnetic flux quantum. For rational $\theta = p/q$, the bulk eigenstates distribute in $q$ energy bands with non-trivial topological properties encoded in their Chern numbers. An example of such band dispersion is shown in Fig.~\ref{fig1}(a) for the specific $\theta=1/4$ case on which we are going to focus throughout this work. \iac{This dispersion is 
\iac{obtained}
by calculating the single-particle eigenmodes of the full two-dimensional Harper-Hofstadter Hamiltonian \eqref{eq:HH} under periodic (resp. open) boundary conditions along the $x$-axis (resp. $y$-axis). Given the translational invariance along $x$, this reduces to a one-dimensional diagonalization problem for each value of $k_x$.}

\iac{In particular, note the chiral edge states 
that appear in the energy gaps between the bands. Their dispersion $\epsilon(k_x)$ is} plotted in Fig.~\ref{fig1}(a) in blue and green lines for the $y=1,n_y$ edges, respectively.
\iac{Two such edge modes exist within each energy gap: they are localized on the opposite $y=1,n_y$ physical edges of the system and propagate with opposite group velocities.} For instance, the edge mode of the negative energy gap with positive group velocity is localized on the $y=1$ side, while the one with negative group velocity on the $y=n_y$ side. The opposite holds for the edge modes in the positive energy gap. 

Some crucial properties of the edge modes localized on the $y=1$ edge are summarized in \iac{Fig.~\ref{fig1}(b)}, namely their effective mass $m_*^{-1} = \partial^2_{k_x}\epsilon(k_x)$
(related to the curvature  of the energy dispersion, blue line) and their overlap with the edge site $y=1$ (red line). The latter is quantified by the edge localization function,
\begin{equation}
\Lambda(k_x) = |\phi_{y=1}(k_x)|^2\,,
\end{equation}
where $\phi_y(k_x)$ is the wavefunction of the edge mode of wavevector $k_x$, normalized according to $\sum_y |\phi_{y}(k_x)|^2 =1$. As expected, the localization is maximum at $k_x$ values for which the edge mode is located around the centre of the energy gap (black dot in \iac{Fig.~\ref{fig1}(a)}).

\subsection{Gain, loss and nonlinear terms}

At the semiclassical level, we can replace the bosonic field operators on each lattice site with 
$c$-number amplitudes and recast the field dynamics in terms of the following equation of motion,
\begin{multline}\label{eq:HHlaser2D}
      i\frac{\partial \psi_{x,y}(t)}{\partial t} =  \Big[ g|\psi_{x,y}|^2  + g_R N_{x,y} 
       + \frac{i}{2}(R N_{x,y} - \gamma) \Big] \psi_{x,y} \\ -J\Big[\psi_{x,y+1} + \psi_{x,y-1}  + e^{-2\pi i \theta y}\psi_{x+1,y} 
       + e^{+2\pi i \theta y}\psi_{x-1,y}\Big].   
\end{multline}
Hopping between neighbouring sites occurs along both $x,y$ directions. In the chosen Landau gauge, the synthetic magnetic field is encapsulated in a $y$-dependent phase of the hopping along $x$.
All lattice sites experience losses at a rate $\gamma$ and the nonlinear refractive index results in an intensity-dependent frequency shift proportional to the nonlinearity coefficient $g$.

\iac{The} gain is provided by a reservoir of incoherent excitations of density $N_{x,y}$ obeying the rate equation,
\begin{equation}\label{eq:res2D}
      \frac{\partial N_{x,y}}{\partial t} = P \delta_{y,1} - (\gamma_R + R|\psi_{x,y}|^2)N_{x,y}
\end{equation}
and describing, e.g., the density of electrons promoted to the conduction band of a semiconductor gain material. 
This reservoir is pumped at a site-dependent rate: as indicated in \eqref{eq:res2D}, we concentrate on the case where \iac{the} pumping is localized on the $y=1$ edge of the lattice and here has a uniform rate $P$. The reservoir decays on a characteristic \iac{timescale} set by $\gamma_R$ and provides stimulated emission into field modes with an efficiency $R$.
The effect of the incoherent excitations on the refractive index, and hence on the resonance frequency of each site, is included by the photon-reservoir interaction term $g_R N_{x,y}$, which is at the origin of the Henry linewidth enhancement factor~\cite{henry1982}. 

A\iac{n e}specially important regime is identified when  the carrier dynamics  is very fast compared to the other timescales of the  device, i.e. for ${\gamma_R}/{\gamma} \gg 1$. In this case we can make use of the adiabatic approximation and set the left-hand side of Eq.~(\ref{eq:res2D}) to zero. 
The carrier density then instantaneously follows the field dynamics according to
\begin{equation}
    N_{x,y} = \frac{P\delta_{y,1}}{\gamma_R + R|\psi_{x,y}|^2}\,.
\end{equation}
In the following, a device satisfying this condition and 
featuring negligible nonlinearities $g,g_R=0$
is referred to as a class-A laser and is described by the following equations of motion for the field amplitude,
\begin{equation}\label{HH2D_ad}
      i\frac{\partial \psi_{x,y}(t)}{\partial t} = (\mathbf{H}\psi)_{x,y}  +  \frac{i}{2}\bigg( \frac{\beta P \delta_{y,1}}{1+\beta |\psi_{x,y}|^2} - \gamma \bigg) \psi_{x,y} 
\end{equation}
where the hopping matrix is such that 
\begin{multline}
    (\mathbf{H}\psi)_{x,y} = -J \big[\psi_{x,y+1} + \psi_{x,y-1}  +  \\ + e^{-2\pi i \theta y}\psi_{x+1,y} + e^{+2\pi i \theta y}\psi_{x-1,y}\big]
\end{multline} 
and the effective saturation parameter \iac{is} $\beta = {R}/{\gamma_R}$.
This simplified model was used in our previous works~\cite{secli2019,amelio2019b}. 

The present work goes beyond this regime and extends the investigation to a more general class of devices,
where the reservoir cannot be adiabatically eliminated and/or significant nonlinearities are present, $g,g_R \neq 0$. \iac{Such a non-adiabatic $\gamma_R<\gamma$ regime is commonly found both in polariton topolaser devices~\cite{klembt2018} and semiconductor laser ones~\cite{bandres2018,longhi2018}: in the former case, this is due to the long recombination rate of the excitons feeding the condensate, while in the latter case it is due to the slow dynamics of carriers in the semiconductor gain medium. On the other hand, optical nonlinearities due to repulsive polariton-polariton and polariton-reservoir interactions $g,g_R>0$ are especially significant in polariton devices where the ensuing blue-shifts $gn$ and $g_Rn_R$ may exceed the loss rate $\gamma$ and even approach the hopping amplitude $J$~\cite{carusotto2013,baboux2018,bobrovska2018dynamical}}.

\section{Steady-state lasing solution}\label{sec:2D_steady_state}

\iac{As usual, the first step in the calculation of the fluctuation dynamics of a laser device consists in characterizing its steady state.} As long as losses overcome the gain, the steady-state  of the device is the electromagnetic vacuum $\psi_{x,y}=0$ and $N_{x,y=1} = P / \gamma_R$. A non-trivial steady state is instead reached when the gain starts exceeding losses. 

The transition between the two regimes defines a threshold value $P_{\textrm{th}}$ for the pump rate $P$, which can be calculated by linearizing the motion equation Eq.~(\ref{eq:HHlaser2D}) for the field $\delta \psi_{x,y}$ around the vacuum solution. Thanks to the translational symmetry of our system along the periodic $x$ direction, it is useful to move to Fourier space along $x$ and, for each $k_x$ value, solve the one-dimensional eigenvalue problem:
%
\begin{multline}\label{seuil_las}
    \omega \, \delta \psi_{k_x,y} =   \frac{i}{2} \Big[ (R-2ig_R) \frac{P}{\gamma_R}  \delta_{y,1}  - \gamma\Big] \delta \psi_{k_x,y} +   \\
    -J  \Big[\delta \psi_{k_x,y+1} + \delta \psi_{k_x,y-1} 
      + 2 \cos(2\pi \theta y + k_x) \delta \psi_{k_x,y}\Big]\,.
\end{multline}
An example of the spectrum of the corresponding $n_y \times n_y$ matrix is shown in Fig.~\ref{fig1}(c) for parameters very close to the lasing threshold. The precise position $P_{\textrm{th}}$ of the threshold can be determined as the point at which the imaginary part of one of the eigenvalues turns positive, meaning that the vacuum solution is no longer dynamically stable.

For $P>P_{\textrm{th}}$ the system departs from the unstable vacuum solution and, under suitable conditions to be discussed better in what follows, it can reach a non-trivial, dynamically stable stationary state displaying a periodic oscillation of the field at some frequency $\omega^{\textrm{Las}}$. \iac{Our numerical study of this dynamics is carried out by solving the evolution equations (\ref{eq:HHlaser2D}-\ref{eq:res2D}) in real time. This is done using a fourth-order Runge-Kutta algorithm, starting from a small random complex amplitude on each $x,y$ site to trigger the instability. In order to characterize the steady-state, we let the evolution run for long enough times (on the order of \iac{$10^6$ times steps}) until a clean steady-state is reached. Typical lattice sizes used in our simulations are $n_x= 64$ and $n_y= 23$ 
with periodic boundary conditions along $x$. Accurate results are obtained with a typical time step on the order of $dt= 0.005/J$.}

\iac{Besides the numerical study, analytical arguments can be used to understand the physics of the steady-state.} Thanks to the translational invariance along the $x$-axis, the monochromatically oscillating steady-state can be \iac{formally} written as:
\begin{eqnarray}
    \psi^{\textrm{ss}}_{x,y}(t) &=&  \psi^0_{x,y} e^{-i\omega^{\textrm{Las}} t}  
    = \psi^0_y e^{-i\omega^{\textrm{Las}} t +i k_{x}^{\textrm{Las}}x}
    \label{eq:steady-state_psi}\\
     N^{\textrm{ss}}_{x,y}(t) &=& N^0 \delta_{y,1} . \label{eq:steady-state_N}
\end{eqnarray}
where the lasing frequency $\omega^{\textrm{Las}}$ is self-consistently chosen by the system dynamics. Unless $P$ is very close to the threshold $P_{\textrm{th}}$,  the lasing wavevector $k_{x}^{\textrm{Las}}$ is randomly selected
within the range of $k_x$ modes for which the vacuum state is dynamically unstable.
This selection is triggered by external noise or by the initial conditions imposed to the field~\cite{secli2019}. The global phase of the oscillating field is randomly selected at each instance of laser operation, but then it stays stable for macroscopically long times: this is a characterizing feature of laser emission and is related to the spontaneous breaking of the global $U(1)$ symmetry of Eq.~(\ref{eq:HHlaser2D}) that occurs above threshold.



In Fig.~\ref{fig1}(c), we illustrate how, for our pumping localized on an edge, the lasing instability is stronger for the chiral edge states localized on the $y=1$ pumped side than for the bulk modes that experience a reduced overlap with the edge. Among the chiral edge states, the ones located around the centre of the energy gap are the most localized in space and thus feel the largest gain. Given the small but significant penetration of the edge states into the lossy bulk, the threshold is pushed at a slightly higher pumping value $P_{\textrm{th}} = 1.142 \gamma \gamma_R/R$ than the one $ P_{\rm th,1} = \gamma \gamma_R/R$ of an isolated resonator. 

As long as the nonlinearites $g,g_R$ can be neglected, the effective gain is equal for the negative and positive frequency edge modes as shown in Fig. \ref{fig1}(c), so the instability occurs with the same probability in each of the chiral edge modes. This is perfectly consistent with our previous numerical study~\cite{secli2019}.
This symmetry that holds for $g,g_R=0$ is due to the extended chiral $\mathcal{C}\mathcal{P}_x\mathcal{T}$ symmetry that our system inherits from the one of the underlying HH model.
Indeed, for the conservative model, we have three discrete symmetries, and we review here their action on a generic eigenstate in the form (\ref{eq:steady-state_psi}).
To begin with, the $\mathcal{P}_y\mathcal{T}$ symmetry  requires that also
\begin{equation}
\left[\mathcal{P}_y\mathcal{T}\psi^{\textrm{ss}}\right]_{x,y}(t) = 
e^{2\pi i \theta (n_y+1)x}
\left( \psi^0_{x,n_y-y+1} \right)^*
e^{-i\omega^{\textrm{Las}} t }
\label{eq:PyTsymmetry}
\end{equation}
is an eigensolution of the same frequency and with momentum $2\pi \theta (n_y+1) - k_x$; importantly, if $\psi^{\textrm{ss}}_{x,y}$ were localized on the $y=1$ side, the transformed state would live on the $y=n_y$ edge. Physically, this symmetry corresponds to a reflection plus time-reversal symmetry of the cyclotron orbits in a Hall bar, connecting edge modes located in the same energy gap and living on different edges of the system.
Analogously, one would define the $\mathcal{P}_x\mathcal{T}$ symmetry
as 
\begin{equation}
\left[\mathcal{P}_x\mathcal{T}\psi^{\textrm{ss}}\right]_{x,y}(t) = 
\left( \psi^0_{n_x-x+1,y} \right)^*
e^{-i\omega^{\textrm{Las}} t }\,.
\label{eq:PxTsymmetry}
\end{equation}
\iac{H}owever, since $\psi^0_y$ is (proportional to) a real vector, this mapping corresponds to simple multiplication by a phase~\footnote{Of course, the different nature of the $\mathcal{P}_x\mathcal{T}$ and $\mathcal{P}_y\mathcal{T}$ transformation in this case originates from the fact that the system is translationally invariant along x}.
Finally, the chiral symmetry 
\begin{equation}
\left[\mathcal{C}\psi^{\textrm{ss}}\right]_{x,y}(t) = (-1)^{x+y} \psi^0_{x,y} e^{+i\omega^{\textrm{Las}} t 
},
\label{eq:Csymmetry}
\end{equation}
defines an eigenstate of opposite frequency $-\omega^{\textrm{Las}}$ and  shifted wavevector $k_x+\pi$~\cite{repellin2019}. This transformation explains the overall symmetric structure of the HH spectrum with respect to the zero-energy point. Evidently, the $\mathcal{C}$ symmetry can only be defined on a lattice and does not exist in a continuum geometry where all Landau levels have positive energy. 

Coming back to the topolaser, the presence of gain and losses breaks the above symmetries; nonetheless, given the steady-state lasing state (\ref{eq:steady-state_psi}), also 
\begin{equation}
\left[\mathcal{C}\mathcal{P}_x\mathcal{T}\psi^{\textrm{ss}}\right]_{x,y}(t) = 
(-1)^{x+y} (\psi^0_{n_x-x+1,y})^* e^{+i\omega^{\textrm{Las}} t}
\label{eq:symmetry}
\end{equation}
is a solution of wavevector $k_x + \pi$, frequency $-\omega^{\textrm{Las}}$ and localized on the same edge. Notice that the lasing state $\psi^0_y$ cannot be taken any more with real entries (see Appendix~\ref{app:gamma/J}), that is why the action of $\mathcal{P}_x\mathcal{T}$ is non-trivial.
For non-vanishing nonlinearities $g,g_R \neq 0$, this $\mathcal{C}\mathcal{P}_x\mathcal{T}$ symmetry breaks down and the lasing states in the two topological energy gaps have different properties. \iac{For instance}, if $g_R > 0$, \iac{($g_R<0$)} the lasing threshold is slightly lower for the positive \iac{(negative)} energy \iac{edge modes}  than  for the negative \iac{(positive)} energy ones~\cite{longhi2018}. 

\section{Collective excitations of class-A topological lasers}
\label{sec:classA}

\begin{figure*}[htbp]
    \centering
    \includegraphics[scale=0.4]{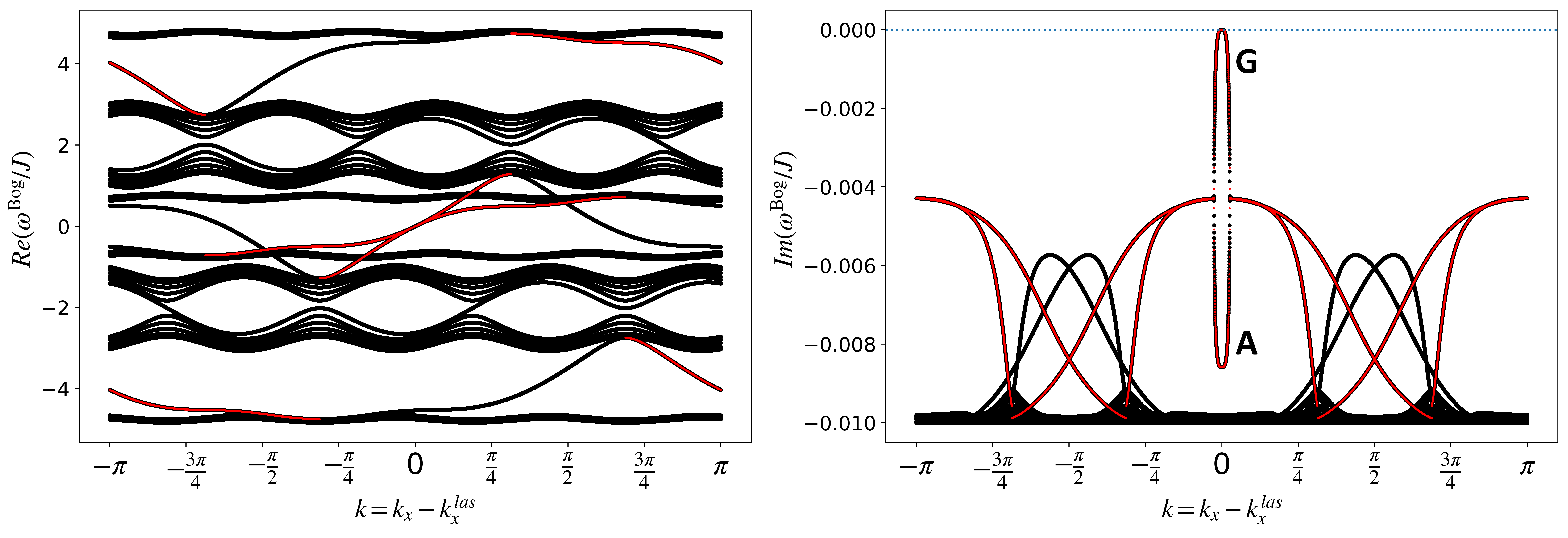}
    \caption{ {Dispersion of the collective excitation modes on top of a class-A topological laser}. The left/right panels show the real/imaginary part of the \iac{Bogoliubov} dispersion \iac{ \iac{for a system of size $n_y=23$} that is lasing on the maximally localized mode at $k_{x}^{\textrm{Las}} = -0.982$ for which gain is strongest.}
    Black (red) dots indicate the results of the full 2D model (1D effective theory). \textbf{G} (resp. \textbf{A}) in the right panel indicate the Goldstone (resp. Amplitude) branches.
    System parameters: $\gamma= 0.02 J$, adiabatic regime \iac{$\gamma_R / \gamma = +\infty$, $P/P_{\rm th,1} = 2$, $g = g_R = 0$.}
    }
    \label{fig:plot_ref}
\end{figure*}

After having identified in the previous Section the steady-state lasing state, we now proceed with the investigation of its collective excitations, namely of the linearized dynamics around the steady-state. This study is the microscopic complement to the statistical study of the coherence properties of the emission~\cite{amelio2019b}, and it provides crucial insight \iac{into} the dynamical stability of the lasing state. 
After presenting the results of a full numerical calculation, we will develop a deeper understanding of the physical features by means of an effective 1D model for the edge state dynamics. In doing \iac{so}, a major focus will be put on the soft Goldstone branch corresponding to the spontaneously broken $U(1)$ symmetry: its dependence on the curvature of the edge mode dispersion and on its  spatial localization on the edge of the physical lattice will be highlighted.
\iac{In this Section, we start our study from the simplest case of class-A devices} displaying a fast reservoir $\gamma_R/\gamma\gg 1$ and no optical nonlinearities $g=g_R=0$, postponing the more general analysis to the next Sec.~\ref{sec:classB}.

\subsection{2D Bogoliubov theory}
\label{sec:2D}

As usual in the Bogoliubov approach~\cite{wouters2007}, the first step of the calculation of the collective modes is to accurately determine the steady-state in the form (\ref{eq:steady-state_psi}). As it was discussed in the previous Section, this can be done by numerically simulating the evolution equation (\ref{HH2D_ad}) \iac{with a suitable Runge-Kutta technique. The lasing frequency $\omega_{x}^{\textrm{Las}}$ and wavevector $k_{x}^{\textrm{Las}}$ are obtained by temporal and spatial Fourier transform of the steady-state field amplitude on the $y=1$ edge.} 

Then, one has to linearize the equations of motion around the steady-state according to the ansatz
\begin{equation}\label{eq:dpsi}
\psi_{x,y}(t) = [\psi^0_y +  \delta \psi_{x,y}(t)]\, e^{-i\omega^{\textrm{Las}} t + i k_{x}^{\textrm{Las}}x}. 
\end{equation}
Thanks to the translational invariance, the collective modes are classified by the $x$ component of the wavevector $k_x$. One can thus switch to Fourier space along $x$ and, \iac{for each value of $k_x$, one can} write a system of linear equations for the corresponding components of the field $\delta \psi_{k_x,y}$ and $\delta \psi_{k_x,y}^*$. 
\iac{The eigenmodes of this linearized evolution are obtained from the $2n_y\times 2n_y$ linear problem determined by}
\begin{equation}\label{eq:HH2D_ad_lin}
      \omega^{\rm Bog}\,\delta \psi_{k,y} = ([\mathbf{H} - \omega^{\textrm{Las}}\mathbf{I}] \,\delta \psi)_{k,y}  + \mathbf{D}_y\, \delta \psi_{k,y} + \mathbf{\tilde{D}}_y \, \delta \psi_{-k,y}^*
\end{equation}
and \iac{the complex conjugate equation.} 
Here, both the wavevector $k= k_x - k_{x}^{\textrm{Las}}$ and the frequency $\omega^{\rm Bog}$ are measured with respect to the lasing ones, $\mathbf{I}$ is the $n_y\times n_y$ identity matrix, and we have defined the short-hands
\begin{eqnarray}
\mathbf{D}_y &=&   \frac{i}{2}\Big(\frac{\beta P \delta_{y,1}}{1+\beta |\psi_{1}^0|^2}  -\frac{\beta^2 P \delta_{y,1} |\psi_1^0|^2}{(1+\beta |\psi_1^0|^2)^2}- \gamma\Big) \\
\mathbf{\tilde{D}}_y &=& -  \frac{i}{2}\frac{\beta^2 P \delta_{y,1} (\psi_1^0)^2}{(1+\beta |\psi_1^0|^2)^2}.
\end{eqnarray}
where $\psi_1^0$ is the component of the steady-state on the edge site $y=1$ that is endowed with gain. Note that the diagonal block $\mathbf{H}$  couples neighbouring sites along $y$, so that in a given $k$ block  all $\delta\psi_{k,:}$ are linearly coupled to $\delta\psi_{-k,:}^*$, where the symbol $:$ indicates a vector with indices $y=1,...,n_y$.
The problem of determining the collective excitation modes then amounts to the numerical diagonalization of a $2 n_{y} \times 2 n_{y}$ Bogoliubov matrix for each $k_{x}$.

A typical example of collective excitation spectrum is displayed by the black dots in Fig.~\ref{fig:plot_ref}. As expected, it displays the characteristic features of non-equilibrium condensates \cite{wouters2007}.
In the real part of the spectrum (left panel), the excitations around the lasing mode (small $\omega^{\rm Bog}$ and $k$) exhibit a zone of adhesion. In this region, the Bogoliubov branch follows the HH edge mode, properly shifted according to $\omega^{\textrm{Las}}$ and $k_{x}^{\textrm{Las}}$. 
In the imaginary part of the spectrum (right panel), we recognize instead the typical splitting between the Goldstone and the amplitude branches, respectively related to phase and intensity fluctuations. As $k \to 0$ the dispersion $\omega^{\rm Bog}_+(k)$ of the Goldstone mode tends to zero in both real and imaginary parts as a consequence of the spontaneous breaking of the $U(1)$ phase symmetry. The amplitude mode has instead a finite negative imaginary part $\mathrm{Im}(\omega^{\rm Bog}_-(k\to 0)) = - \Gamma$ corresponding to the relaxation rate of intensity fluctuations $\Gamma= \gamma(1-P_{\textrm{th}}/P)<\gamma$ (see the 1D model of the next paragraph for the derivation of this formula): for $P\gg P_{\textrm{th}}$ above threshold, $\Gamma$ recovers the bare decay rate $\gamma$, closer to the threshold it is smaller than $\gamma$, and tends to zero right above the threshold $P_{\textrm{th}}$.

One of the peculiarities of our HH laser is the presence of another edge mode with opposite chirality, living on the same edge in the other energy gap at opposite wavevector. In the excitation spectrum, this additional edge mode corresponds to the maximum of the imaginary part around $k \simeq \pm \pi$ .
Since it is also localized on the same edge, this opposite chirality mode also benefits of gain and thus displays a slower decay rate $\Gamma$ than the bulk modes. These latter have in fact a negligible overlap with the gain material and thus decay at the bare loss rate $\gamma$.

\iac{Altogether, these numerical calculations confirm the dynamical stability of topolaser devices in the regime where the gain medium adiabatically follows the field dynamics and no other optical nonlinearity is present besides gain saturation. Getting analytical insight \iac{into} the physics underlying \iac{these} numerical results will be the subject of the next Subsection.}

\subsection{Effective 1D model}
\label{sec:1D}

We now proceed to develop an effective 1D model that is able to provide analytical insight into the collective excitation spectra numerically calculated in the previous Subsection using the full 2D theory. 
This method relies on the assumption that the lasing wavefunction closely follows the corresponding eigenstate of the underlying conservative HH model. First, notice that \iac{the} translational invariance along $x$ allows writ\iac{ing} the steady-state as a plane wave of quasi-momentum $k^{\rm las}_x$ also in the nonlinear case. Then, one formulates the ansatz 
\begin{equation}\label{eq:ss_ansatz}
    \psi^{\textrm{ss}}_{x,y}(t) \simeq \psi^{\textrm{ss}} e^{i k_x^{\textrm{Las}}x} \phi_y(k_x^{\textrm{Las}})\,e^{-i\omega^{\textrm{Las}} t} 
\end{equation}
where $\phi_y(k_x)$ is the transverse wavefunction of the edge mode at wavevector $k_x$. This writing is expected to be accurate in the $\gamma \ll J$ limit where the band gap is much wider than the frequency scale of the dynamics. A brief discussion of the first corrections in $\gamma/J$ is given in the Appendix~\ref{app:gamma/J}.

We will also consider small fluctuations on top of this solution, in the form 
\begin{equation}
    \psi_{x,y}(t)=\int\!\frac{dk}{2\pi}\,e^{i(k_x^{\textrm{Las}} + k) x}\,\tilde{\psi}(k,t)\,\phi_y(k+k_x^{\textrm{Las}})\,e^{-i\omega^{\textrm{Las}} t} 
    \label{eq:1D_ansatz}
\end{equation}
where $\tilde{\psi}(k,t) =  2\pi\,\delta(k)\,\psi^{\textrm{ss}}  + \delta\tilde{\psi}(k,t)$, with $\delta\tilde{\psi}(k,t)$ small.
The goal of this subsection is to determine an equation of motion for the 1D wavefunction 
\begin{equation}
\psi(x,t) \equiv \int\!\frac{dk}{2\pi}\,e^{i(k_x^{\textrm{Las}} + k) x}\,\tilde{\psi}(k,t)\,e^{-i\omega^{\textrm{Las}} t}.
\end{equation}

As a first step, we are now going to determine the amplitude $\psi^{\textrm{ss}}$ of the plane wave wavefunction at the steady-state. To this purpose, we inject the ansatz (\ref{eq:ss_ansatz}) into the evolution equation, and then we overlap
the result with $\phi_y(k_x^{\textrm{Las}})$
\footnote{Obviously, in the general case $\gamma \sim J$ the non-uniform gain couples the edge mode to the bulk modes of equal $k_x$, so that it is no longer sufficient to overlap with the edge modes only. This coupling becomes negligible for $\gamma \ll J$.}. This gives
\begin{multline}
0 = 
\epsilon(k^{\textrm{Las}}_x) - \omega^{\textrm{Las}} + \frac{i}{2} \left[ \frac{\beta P \Lambda(k^{\textrm{Las}}_x)}{1+\beta \, \Lambda(k^{\textrm{Las}}_x) 
\, |\psi^{\textrm{ss}}|^2}   - \gamma  \right] 
\end{multline}
where 
$\epsilon(k_x)$ is the edge mode dispersion and $\Lambda(k_x) = |\phi_{y=1}(k_x)|^2$ is the edge localization function that quantifies the overlap of the HH edge states with the $y=1$ edge.
Splitting the real and imaginary parts of this equation, we obtain
\begin{eqnarray}
\omega^{\textrm{Las}} &=& \epsilon(k^{\textrm{Las}}_x) \label{eq:ss_field_mu}\\
|\psi^{\textrm{ss}}|^2 &=& \frac{1}{\beta \Lambda(k^{\textrm{Las}}_x) 
}\,\left(P/P_{\mathrm{th}} - 1\right)\,.
\label{eq:ss_field_intensity}
\end{eqnarray}
The lasing frequency is set by the bare dispersion of the edge state, whereas the lasing threshold 
\begin{equation}
    P_{\mathrm{th}} =  \frac{\gamma}{{\beta\Lambda(k_x^{\textrm{Las}})}}
\end{equation} 
depends on the overlap of the mode with the gain region via the $\Lambda(k_x)$ localization factor.  As one can see on the red lines in Fig.\ref{fig1} (b), for the most localized modes, this factor can reach values close to unity.

As the next step, we try to write the equation of motion for $\psi(x,t)$ 
below threshold. Since the equation of motion in this regime is linear, the different Fourier components  decouple and one can write
\begin{equation}
      i\partial_t \psi(x,t)  =\left[ 
    \epsilon(\hat{k}_x)   + \frac{i}{2}\left( \Lambda(\hat{k}_x)  \beta P - \gamma \right) \right] \psi(x,t)\,
    \label{eq:dynamics_lin}
\end{equation}
where $\hat{k}_x=-i\partial_x$ is the usual momentum operator and with the only assumption that the coupling to bulk modes is negligible. The conservative part of the dynamics follows the HH edge dispersion and the effective gain strength has a $k$-dependence given by the edge localization function $\Lambda$: the stronger the localization, the stronger the effective gain.
This equation being linear, its collective excitation modes are trivially given by 
\begin{equation}
\omega^{\rm Bog}(k)= \epsilon(k_x) - \omega^{\textrm{Las}}  + \frac{i}{2}\Big( \Lambda(k_x)  \beta P - \gamma\Big)\,.
\end{equation}
where $k$ is the momentum with respect to the lasing one, $k_x \equiv k_x^{\textrm{Las}}+k$.
This effective 1D prediction for the dispersion around the vacuum state is plotted as a red line in Fig.~\ref{fig1}(c): as long as one focuses on the edge modes, it excellently recovers the full 2D calculation shown by the black lines.
In particular, the 1D model provides a reliable prediction for the mode with the strongest gain, which is going to lase first. Of course, the full 2D calculation also includes the bulk bands that are not captured by the 1D model: however, given their smaller overlap with gain, the imaginary part of their frequency is much larger and negative. 

While the linear dynamical equation \eqref{eq:dynamics_lin} is exact in the $\gamma/J\to 0$ limit, extending it to the nonlinear regime where gain saturation is important is made non-trivial by the simultaneous $x$- and $k_x$-dependence of the gain term: gain saturation is in fact a spatially local effect, while the $k_x$ dependence of gain via the edge localization function is a momentum-space effect. In the spirit of our previous discussion, 
one can generalize Eq.~(\ref{eq:dynamics_lin}) and write down
\begin{equation}\label{eq:1D_wire}
      i\partial_t \psi(x) = \left[ 
    \epsilon(\hat{k}_x)  + \frac{i}{2}\bigg( \Lambda(\hat{k}_x)  \frac{\beta P }{1+\beta |\psi(x)|^2} - \gamma \bigg) \right] \psi(x) . 
\end{equation}
Notice that the edge localization and the saturation terms do not commute and the order 
has been chosen to get consistent results with the 2D theory.

While we provide no rigorous derivation of this formula~\footnote{An attempt of derivation would involve the Bogoliubov ansatz
$\delta \psi_{x,y} = u_k e^{ikx} \phi_y(k) +  v_k e^{-ikx} \phi_y(-k)$; the corresponding Bogoliubov problem would not be closed since in general $\phi_y(k) \neq \phi_y(-k)$.},
we show that the Bogoliubov edge eigenenergies are accurately recovered and conveniently interpreted within this approach. 
Indeed, the Bogoliubov equations for Eq.~(\ref{eq:1D_wire}) can be cast in the usual $2 \times 2$ matrix form
\begin{widetext}
\begin{equation}\label{eq:bogo_mat_1D_ad}
    \omega^{\rm Bog}(k) \begin{pmatrix}
    u_k \\ v_k
\end{pmatrix}     
    = 
        \begin{bmatrix} e(k) + \frac{i\gamma}{2} \left( \lambda(k)-1 \right) - \frac{i}{2}\Gamma \lambda(k) &  -  \frac{i}{2}\Gamma \lambda(k)\\ -  \frac{i}{2}\Gamma \lambda(-k) &  -e(-k) +  \frac{i\gamma}{2}  \left( \lambda(-k)-1 \right) -  \frac{i}{2}\Gamma \lambda(-k)\end{bmatrix}
    \begin{pmatrix}
    u_k \\ v_k
\end{pmatrix} 
\end{equation}
\end{widetext}
where we have defined the shorthands $e(k)=\epsilon(k_x^{\textrm{Las}}+k)-\omega^{\textrm{Las}}$ and $\lambda(k)=\Lambda(k_x^{\textrm{Las}}+k)/\Lambda(k_x^{\textrm{Las}})$.
Expanding the HH edge state dispersion $\epsilon(k) - \omega^{\textrm{Las}} \simeq v_g k + \frac{k^2}{2m_*}  $, it is immediate to see that the group velocity (as well as the higher odd terms of the dispersion) gives a diagonal term that contributes as a constant to the Bogoliubov dispersion. The $\lambda(k)$ coefficient is of geometric nature and accounts for the $k$-dependence of the edge mode localization. The Bogoliubov spectrum  that results from the diagonalization of this matrix 
consists of two branches $\omega^{\rm Bog}_{\pm}(k)$
and is plotted as red lines in Fig.~\ref{fig:plot_ref}. The agreement with the full 2D numerical calculation is excellent: both the Goldstone  and amplitude  branches are quantitatively recovered by $\omega^{\rm Bog}_{+}(k)$ and $\omega^{\rm Bog}_{-}(k)$ respectively, as well as the dispersion of the edge mode with opposite chirality. Of course, the bulk bands are not included in the 1D model.
 
\begin{figure}[htbp]
    \centering
    \includegraphics[width=\columnwidth]{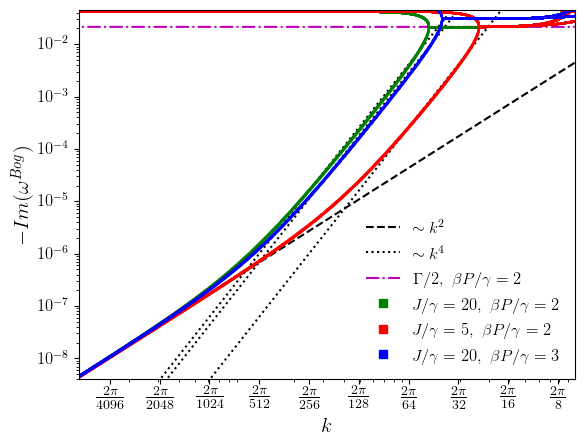}
    \caption{ {Scaling of  the Goldstone branch at small k}. The red, green and blue lines show the Goldstone and the amplitude branches, calculated with the full 2D model for different hopping strengths and pumping parameters, as detailed in the legend. The thick black dashed line is a quadratic fit of $ -\mathrm{Im}(\omega^{\rm Bog}(k))$, on top of which all the three dispersion relations collapse at small $k$. The thin black dotted lines correspond to the $(k^2/2m_*)^2/{\Gamma}$ scaling predicted by the 1D model. The magenta dash-dotted line indicates the relaxation rate at which Goldstone and amplitude branches stick for \iac{${P}/P_{\rm th,1}=2$.}
    \iac{Other parameters: $k_{x}^{\textrm{Las}} = -0.982$, adiabatic \iac{regime} $\gamma_R/\gamma=\infty$, $g = g_R = 0$.}
    }
    \label{fig:Bogo_scaling}
\end{figure}

Among all Bogoliubov modes, the ones with the slowest relaxation rate play a very important role in determining the long-distance, long-time behaviour of the spatio-temporal coherence properties of the laser emission~\citep{amelio2019b}. 
If the localization of the edge mode was uniform in $k$, $\lambda(k)=1$ (as assumed, for instance, in \cite{zapletal2020}), Eq.~(\ref{eq:bogo_mat_1D_ad}) would predict a quartic $\sim k^4$ behaviour of the decay rate of the Goldstone branch at small $k$, proportional to the curvature of the edge mode or, equivalently, to the inverse of the effective mass $1/m_*$, $\mathrm{Im}(\omega^{\rm Bog}_+(k)) \simeq - {(k^2/2m_*)^2}/{\Gamma}$. 
Upon closer inspection of our theory, however, one notices that the $k$-dependence of the localization function entails a quadratic behaviour \begin{equation}
    \mathrm{Im}(\omega^{\rm Bog}_+(k)) \simeq - \frac{\gamma}{2}\frac{\lambda''(0)}{2}k^2
\end{equation} for $k\to 0$, which has  geometric origin and is independent of $J/\gamma$ and, thus, of the effective mass $m_*$.


All these non-trivial predictions of the 1D model are well confirmed by the exact 2D dispersion, plotted on a magnified scale in  Fig.~\ref{fig:Bogo_scaling} (though, there is some quantitative discrepancy in the coefficient of $k^2$, not shown).
For a few different values of $J/\gamma$ and of the pumping $P$, the three different dispersions fall on the same curve at very small $k$.
At intermediate $k$, the curvature of the HH edge mode starts playing a crucial role and, for sufficiently large $J/\gamma$, the imaginary part matches the $\sim (k^2/2m_*)^2/{\Gamma}$ behaviour. 
\iac{For even larger $k$, the imaginary parts of the Goldstone and the amplitude branches stick at a value $-\Gamma/2$ determined by the relaxation rate of small wavevector density fluctuations $\Gamma = \gamma(1-P_{\textrm{th}}/P)$.}

These results show how the effective 1D model is able to reproduce the qualitative features of the edge Bogoliubov modes and, in particular, to explain the $\sim k^2$ behaviour that is crucial for the topological enhancement of coherence in large   lattices~\cite{amelio2019b}.
As a great advantage, the effective one-dimensional theory introduced here for the specific case of a HH model can be straightforwardly adapted to topological lasers built on top of different topological models, e.g. as the Haldane model considered in~\cite{zapletal2020}, by just plugging in the suitable forms of the edge mode dispersion $\epsilon(k_x)$ and of the localization function $\Lambda(k_x)$.

\section{Dynamical stability of general topological laser devices}
\label{sec:classB}

In the previous Sections, we have studied the dispersion of the collective excitations in \iac{the idealized} case of a fast gain medium and no optical nonlinearity besides gain saturation. In particular, we have shown that no dynamical instability occurs in this regime and the only slow dynamics is the one of the Goldstone mode, intrinsically related to the U(1) symmetry breaking mechanism of laser operation.
In this Section, we extend our study to a wider class of devices where the carrier dynamics in the gain medium has a slower timescale than the bare dynamics of the lasing mode and/or the lattice resonators and/or the gain medium display an intensity-dependent refractive index. Our investigation extends the pioneering analysis carried out in~\cite{longhi2018} and provides physical insight \iac{into} the \iac{different microscopic processes underlying the instabilities predicted there}. 

\iac{As in the previous Sections, our analysis will } make a combined use of full two-dimensional calculations, based on a linearized theory that \iac{now} includes the reservoir dynamics into the Bogoliubov formalism as summarized in Appendix~\ref{app:Bogo2D}, and of an effective one-dimensional theory that generalizes Eq.~(\ref{eq:1D_wire}) to the more complex configurations under investigation here. \iac{This one-dimensional theory is based on the following pair of evolution equations for the lasing field and the reservoir density}
\begin{eqnarray}
       i\partial_t \psi(x) &=& \left[ 
    \epsilon(\hat{k}_x) +g|\psi(x)|^2+\right. \label{eq:1D_wire_B} \\
    &+& \left.\frac{i}{2}\bigg( \Lambda(\hat{k}_x)  (R-2ig_R) N(x) - \gamma \bigg) \right] \psi(x) , \nonumber \\
      \frac{\partial N(x)}{\partial t} &=& P  - (\gamma_R + R|\psi(x)|^2)N(x)\,.\label{eq:res2D_1D}
\end{eqnarray}


\begin{figure*}[htbp]
    \centering
    \includegraphics[width=\textwidth]{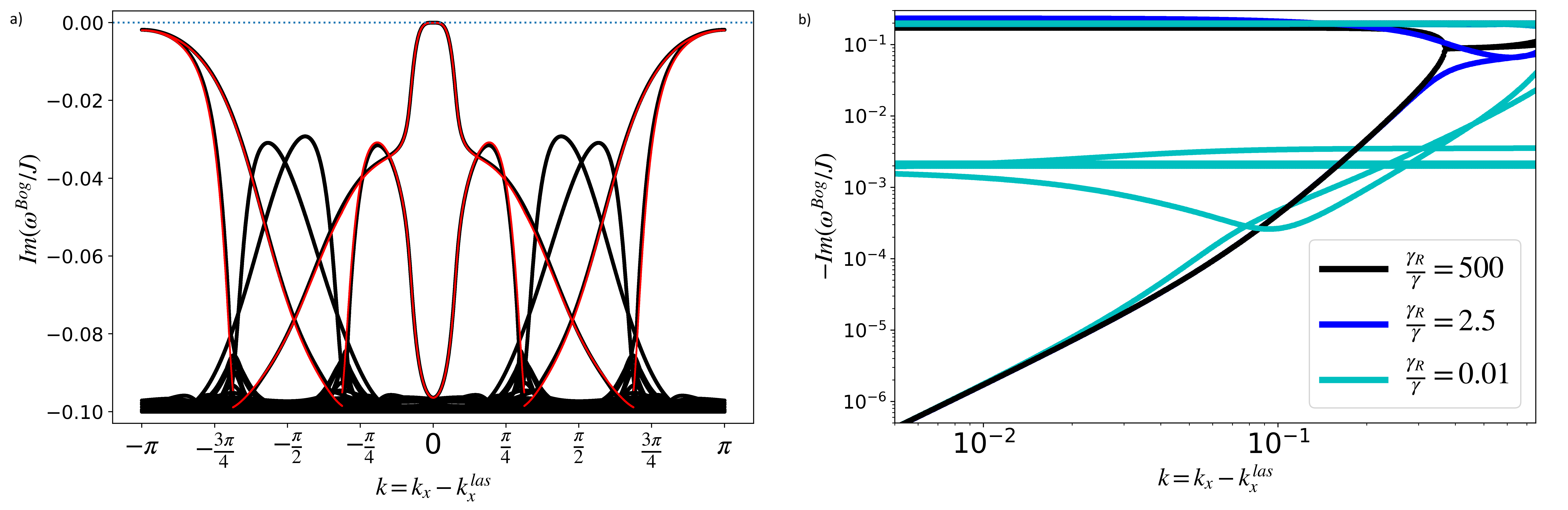}
    \caption{
    Left panel: imaginary part of the elementary excitation spectrum with a slow reservoir $\gamma_R/\gamma = 2.5$. Black (resp. red) lines stand for results computed with the full 2D (resp. 1D effective) model. Right panel: comparison of the small $k$ scaling of the Goldstone branch for different values of the reservoir speed, calculated with the 2D theory. 
    Parameters: $\gamma= 0.2 J$, \iac{$P/P_{\rm th,1} = 2$}, $g = g_R = 0$, $k_{x}^{\textrm{Las}} = -0.982$, \iac{$n_y=23$.}
    }
    \label{fig_slow_res}
\end{figure*}

\subsection{Slow carrier dynamics}\label{sec:slow_res}

In this \iac{first} subsection, we focus on the effect of a slow carrier dynamics \iac{$\gamma_R \lesssim \gamma$} on the dynamical stability of monochromatic laser operation. For simplicity, we assume that no other nonlinearity is present besides gain saturation, that is, we set the intensity-dependent refractive index to zero, $g=g_R=0$.

The imaginary part of the dispersion of the collective excitation modes is shown in Fig.~\ref{fig_slow_res}(a) for the case of a moderately slow carrier dynamics with $\gamma_R/\gamma$ of order $1$.
The black dots show the result of a full 2D calculation of the collective excitation modes. Quite interestingly, also in this case the 1D model (red lines) is able to recover the full 2D calculation in a remarkably accurate way.
While for relatively large wavevectors $k$, the overall shape of the Goldstone and amplitude branches is deeply changed due to the hybridization between edge and reservoir modes, the dispersion of the slowest excitation modes at low $k$ maintains the same structure. 

This physics is displayed in better detail in \iac{Fig. \ref{fig_slow_res}(b).}
Independently of the value of $\gamma_R$, down to the smallest values of $\gamma_R/\gamma$,
for sufficiently small $k$ the Goldstone branch turns slow compared to the reservoir dynamics: in this window of small $k$ values, the reservoir can thus be adiabatically eliminated, and the dispersion recovers a $\gamma_R$- and $m_*$-independent, quadratic $\sim k^2$ dependence as discussed in the previous Section. Of course, the amplitude branch and the higher-$k$ part of the spectrum depend instead strongly on $\gamma_R$.

As a further consequence of the reduced value of $\gamma_R/\gamma$, in \iac{Fig. \ref{fig_slow_res}(a)} one can see how the \iac{counter-propagating edge mode of wavevector $k=\pi$ with opposite chirality} gets  closer to the instability threshold. In the adiabatic limit discussed in the previous Section, we saw that its imaginary part was (in absolute value) equal to $\Gamma/2$, \iac{that is} $\Gamma/J= 0.0858$ for the parameters of \iac{Fig. \ref{fig_slow_res}(a)}. This value is way larger than the numerically calculated value $-\mathrm{Im}[\omega^{\rm Bog}(\pi)]\simeq 0.002 J$. 
Physically, this reduced value can be understood in terms of the high frequency $\approx 2\omega^{\textrm{Las}}$ at which the lasing edge mode (of frequency $\omega^{\textrm{Las}}$ \iac{with respect to the bare frequency of the sites}) beats with the counter-propagating mode (of frequency approximately $-\omega^{\textrm{Las}}$), way higher than the carrier relaxation rate $\gamma_R$. As a result, the fast oscillating interferences are ineffective in quenching the effective gain experienced by the counter-propagating mode. Using the linearized form of the one-dimensional equation of motion, one sees that the imaginary part of the counter-propagating excitations scales as
\begin{equation}
    \mathrm{Im}(\omega^{\rm Bog}(\pi)) = - \frac{\alpha (1+\alpha)}{2} \left(\frac{\gamma_R}{2\omega^{\textrm{Las}}}\right)^2\,\gamma
\label{eq:counter_gamma}
\end{equation}
\iac{with the} shorthand $\alpha=P/P_{\textrm{th}}-1$. \iac{Since $\omega^{\rm Las}$ is typically of order $J$, the relaxation rate \eqref{eq:counter_gamma} of the counter-propagating mode turns out much smaller not only than the bare cavity decay $\gamma$, but also of the carrier one $\gamma_R$.} 

\iac{Even though from a purely mathematical perspective this extremely slow decay time is not harmful to the dynamical stability of topolaser operation, in practice it} may be problematic for applications, since it \iac{may dramatically slow} down the process of selecting one chiral edge mode over the other. \iac{In the transient, the simultaneous presence of oscillations in both chiral modes results in a multimode emission or, from a different point of view, a fast modulation of the laser amplitude at a frequency $2\omega^{\rm Las}$. Beyond this, in Sec.~\ref{sec:general}, we will see how the small value of the imaginary part of the counter-propagating mode makes it susceptible of getting dynamical unstable once nonlinearities are included in the model.}

\begin{figure}[htbp]
    \centering
    \includegraphics[width=\columnwidth]{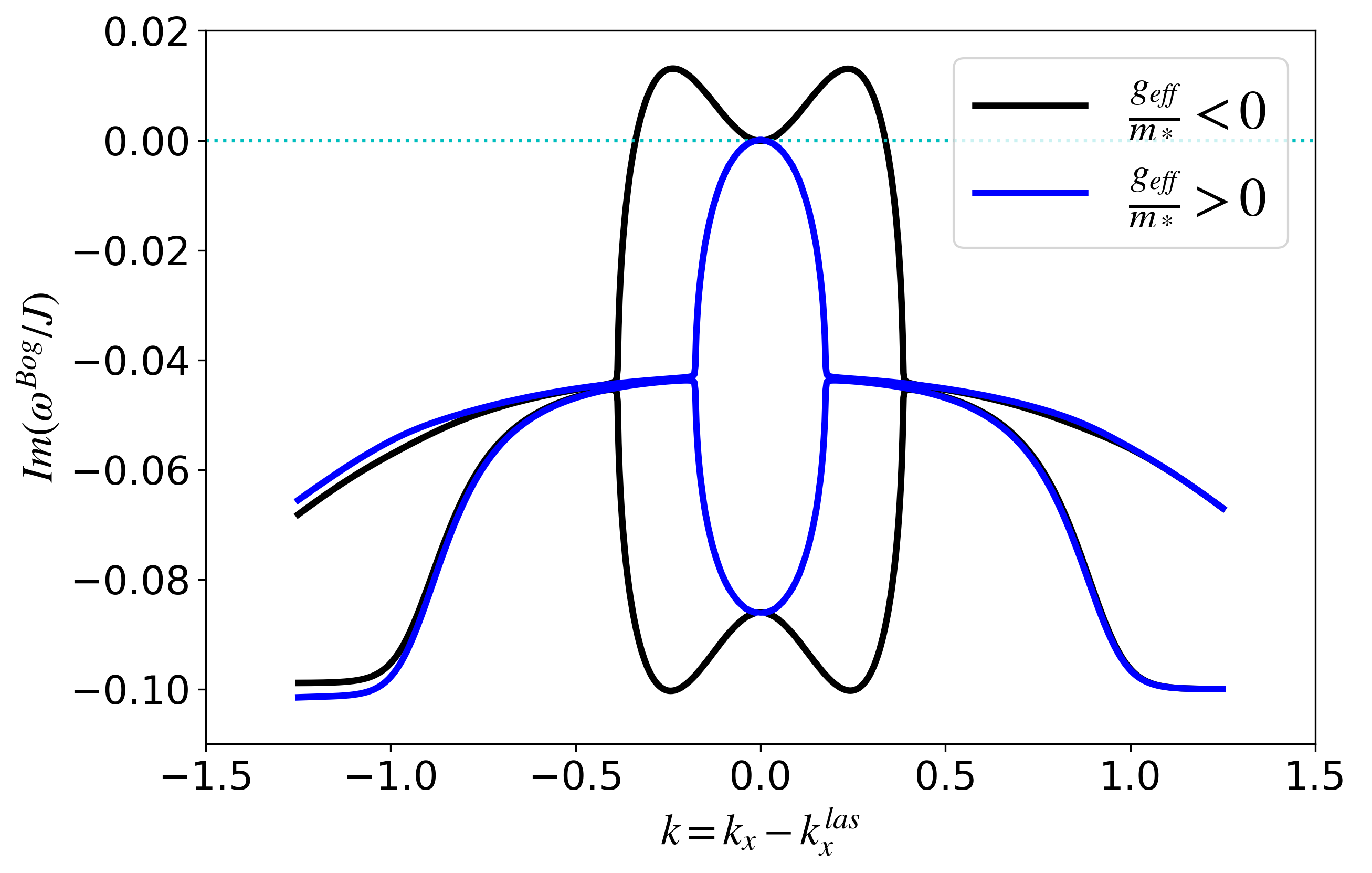}
    \caption{{Imaginary part of the elementary excitation spectrum with small nonlinearities $g|{\psi^{0}_1}|^2, g_R N^{0}\ll J$ in the adiabatic regime, calculated with the 1D effective theory}. Black (resp. blue) corresponds to unstable $g_{\textrm{eff}}/m_* <0$ (resp. stable $g_{\textrm{eff}}/m_* >0$) regime.   
    Parameters: $k_{x}^{\textrm{Las}} = -0.954$, $\gamma= 0.2 J$, \iac{$P/P_{\rm th,1} = 2$}, $g/\beta=0.05 J$, $g_R = 0$. For these parameters, $g \Lambda(k^{\textrm{Las}}) |\psi^{ss}_{x}|^2= 0.037 J \ll J$.
    }
    \label{fig:modul_insta}
\end{figure}

\subsection{Optical nonlinearity}

\label{sec:nonl}

We now investigate the effect on the dynamical stability of a \iac{relatively} small optical nonlinearity such that $g|\psi^{0}_1|^2, g_R 
N^{0}
\ll J$. Under this condition, the transverse profile of the lasing \iac{edge} mode remains similar in shape to the eigenstates of the underlying conservative and linear HH model. Since much of the interesting physics occurs on the slow Goldstone mode, we focus our investigation on this branch and, \iac{starting from the fast reservoir $\gamma_R \gg \gamma$} limit, we adiabatically eliminate the carrier dynamics. 

The dispersion of the Goldstone and amplitude modes in this regime are shown in Fig.~\ref{fig:modul_insta}. The amplitude mode is always stable with an imaginary part $-\Gamma$ at $k=0$. The stability of the Goldstone mode depends instead on the sign of the nonlinearity. This effect \iac{can be understood} in terms of the standard theory of modulational instability in nonlinear optical media~\cite{agrawal1995nonlinear} or \iac{in dilute Bose-Einstein} condensates~\cite{BECbook}. As in~\cite{baboux2018}, once the carrier dynamics has been adiabatically eliminated\iac{, we can define an effective nonlinearity} as 
\begin{equation}
    g_{\textrm{eff}}=g-g_R (\gamma/\gamma_R)(P_{\textrm{th}}/P).
    \label{eq:g_eff}
\end{equation} 
If the effective nonlinearity has the same sign as the curvature of the bare edge mode dispersion (or, equivalently, of the effective mass), the imaginary part of the collective mode dispersion is negative, and the system is dynamically stable. Conversely, if the two quantities have opposite signs, the imaginary part turns positive in a window of wavevectors surrounding $k=0$, signalling dynamical instability of the spatially uniform solution. 

Such behaviour can be understood in \iac{the framework} of the one-dimensional theory \eqref{eq:1D_wire} \iac{by} including an effective interaction term proportional to $g_{\textrm{eff}}\,|\psi(x)|^2$. Setting for simplicity 
$\lambda(k)=1$, one finds the simple form
\begin{equation}
    \omega^{\rm Bog}_+(k) \simeq v_g k -i   \frac{g_{\textrm{eff}}\,|\psi^{\rm ss}|^2} {  m_* \Gamma} k^2 + \mathcal{O}(k^3)\,
\end{equation}
from which it is easy to see how the sign of the imaginary part at low-$k$ is determined by the sign of $g_{\textrm{eff}}/m_*$. \iac{As usual, the observable consequence of this kind of instabilities is a slow spatial modulation of the field amplitude with a wavevector roughly determined by the position of the maximum of the imaginary part and, eventually, its possible break-up into a train of solitons.}

While the sign of the nonlinearity is typically fixed by the material properties of the device, the effective mass of the HH dispersion has opposite sign for edge modes in either the positive and negative frequency energy gap, as illustrated in Fig.~\ref{fig1}(b). \iac{This has the remarkable consequence that, for a given sign of the nonlinearity, topological lasing turns out to be unstable in one of the two frequency gaps and dynamically stable in the other gap. Here, interestingly, the dynamical stability is reinforced by the nonlinearity as signalled by the larger value of the $k^2$ coefficient of the Goldstone mode.  
In mathematical terms, the different behavior of the edge states in the two topological gaps} can be understood as a consequence of the breaking of the chiral symmetry Eq.~(\ref{eq:symmetry}) by the optical nonlinearity. 



\begin{figure*}[htbp]
    \centering
    \includegraphics[width=\textwidth]{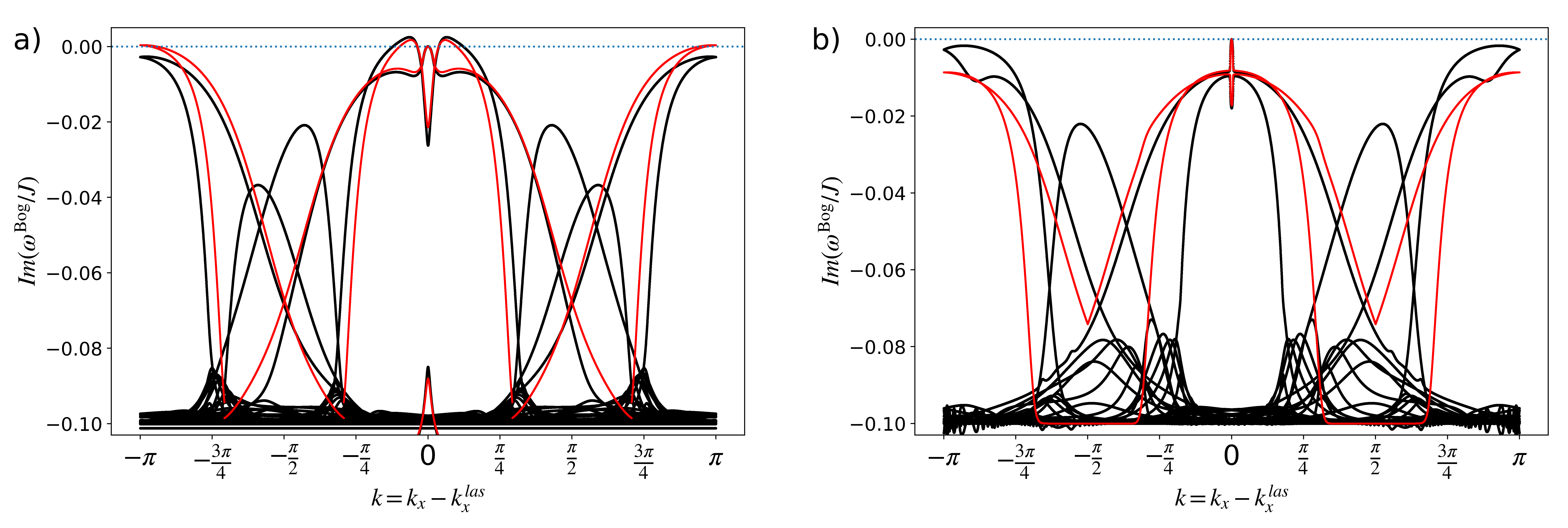}
    \caption{
    Panel (a): imaginary part of the elementary excitation spectrum  with slow reservoir relaxation rate $\gamma_R = 0.5 \gamma$ \iac{for sizable nonlinear interactions} $g_R/R= -1.5 $ between the lasing mode and the carriers \iac{and $g=0$}. 
    Panel (b): imaginary part of the elementary excitation spectrum in the adiabatic regime for a strong repulsive optical nonlinearity $g/\beta= 3.5 J$ and $g_R=0$ \iac{and a pump strength $P/P_{\rm th,1}= 1.25$.
    For these parameters, the lasing frequency \iac{is} $\omega^{\textrm{Las}} - \epsilon(k_x^{\textrm{Las}}) \approx g \Lambda(k^{\textrm{Las}}) |\psi^{ss}|^2 \approx 0.21 J$. Lasing occurs at $k_{x}^{\rm Las}=-0.919$ in panel (a) and $k_{x}^{\rm Las}=-0.982$ in panel (b); the black lines are the result of numerical 2D calculations, while the red lines are the prediction of the effective 1D model. In both panels $\gamma= 0.2J$. $n_y =31$ for panel (a) and $23$ for panel (b)}.
    }
    \label{fig:nonlinear}
\end{figure*}

\subsection{Interplay of nonlinearity and slow carrier dynamics}
\label{sec:interplay}

The pioneering work~\cite{longhi2018} has predicted the occurrence of unstable regimes when a slow carrier relaxation rate $\gamma_R\lesssim \gamma$ is combined with a \iac{sizable nonlinear refractive index due to the carriers in the gain material, a quantity proportional to $g_R n_R$ in our model.}

\iac{The imaginary part of the elementary excitation spectrum in such a regime is plotted in Fig.~\ref{fig:nonlinear}(a) for the case of lasing into a positive mass edge mode in the presence of a relatively slow reservoir $\gamma_R/\gamma=0.5$ and a negative carrier-induced nonlinearity $g_R<0$, $g=0$. 
While for very small $k$ the positive effective nonlinearity \eqref{eq:g_eff} conspires with the positive effective mass to give a stable Bogoliubov mode, a marked instability occurs at slightly larger $k$ (around $|k|\sim 0.2$ for the parameters in the figure) due to the hybridization of the laser and the carrier dynamics. Also in this case, the observable consequence of the instability is the appearance of a spatial modulation of the field amplitude, with an oscillation wavevector roughly determined by the position of the maximum of the imaginary part.
Since this physics has a predominantly one-dimensional character, it is well captured by the one-dimensional theory of (\ref{eq:1D_wire_B}-\ref{eq:res2D_1D}), as displayed by the red lines in \iac{Fig.~\ref{fig:nonlinear}(a)}.}
\iac{On the other hand,} the origin of the visible discrepancies at larger $k$ can be attributed to the distortion of the transverse field profile from the one of the bare HH modes induced by the optical nonlinearity.
In particular, \iac{for this choice of parameters} the modes with reverse chirality \iac{turn out feeling} a lower gain than predicted by the 1D model, 
\iac{and are therefore less prone to dynamical instabilities.}

\iac{Remarkably, very similar behaviours were studied in the context of polariton condensates~\cite{wouters2007,bobrovska2018dynamical,baboux2018} and physically understood in the terms of their interaction with the reservoir of incoherent excitations feeding them: for a positive effective mass, \iac{positive interactions with the reservoir $g_R>0$} correspond to repulsive interactions between the condensate and the slow incoherent reservoir. Therefore, a local increase of the reservoir density pushes the condensate particles away creating a local depletion of their density. This depletion, in turn, reduces the spatial hole-burning effect and leads to a further increase of the reservoir density.  
This provides a positive feedback and makes the initial fluctuation to further grow in time.
\iac{This process explains why the lowest-$k$ Bogoliubov modes are unstable in the $m^*>0$ and $g_R>0$ case. An opposite behaviour is found in the negative $g_R<0$ case considered in Sec.\ref{sec:interplay} or in the negative mass $m^*<0$ case considered in~\cite{baboux2018}: in both these cases, the interactions with the carriers tend to further stabilize laser operation against the slowest Bogoliubov modes. The situation of course changes for modes at increasing $k$, whose frequency no longer satisfies the adiabaticity condition underlying \eqref{eq:g_eff}: for these, one can no longer restrict to the effective interaction $g_{\rm eff}$ and an instability indeed appears, as displayed in Fig.~\ref{fig:nonlinear}(a).}} 


\iac{Quite interestingly, as it was observed in the polariton context~\cite{baboux2018}, this finite-wavevector instability is tamed in the presence of a strong enough wavevector-dependence of \iac{the} gain away from the highest-gain condensate mode. In the case of topolasers, this wavevector dependence can be reinforced with a suitable design of the underlying topological lattice, e.g. via the edge localization function $\Lambda(k_x)$.}



\subsection{Remarks on the general case}
\label{sec:general}

\iac{In the previous subsections, we have seen instabilities occurring either in the neighborhood of the lasing mode at $k\sim 0$ or on modes with opposite chirality at $k\sim \pi$. This appears to be a general feature and is confirmed by the calculations for strong optical nonlinearities.} 

\iac{An example of such calculation is displayed in Fig.~\ref{fig:nonlinear}(b) but the general trend remains the same for other choices of parameters. Here, the nonlinear frequency shift $gn$ 
is positive and comparable to the hopping rate $J$ and induce a significant distortion on the edge modes. Still, the imaginary part remains relatively large and negative throughout all the Brillouin zone except for the regions around $k=0,\pi$, where there exist edge modes well localized on the edge with a strong overlap with the gain medium. All other modes mostly reside in the bulk of the lattice and thus feel a reduced gain, which ensures their dynamical stability.

In the $k\sim 0$ region, the positive mass and the positive nonlinear shift conspire to stabilize the Bogoliubov mode via the same physical mechanism discussed in Sec.~\ref{sec:nonl} in terms of the effective one-dimensional theory. In the $k\sim \pi$ region, instead, the dynamical stability/instability of the excitation modes depends in a less straightforward way on the system parameters: in the specific case of Fig.~\ref{fig:nonlinear}(b), for instance, single-mode lasing is stable, but easily turns unstable as soon as the carrier relaxation rate is decreased or the pump strength is increased. The experimental signature of such instability would be the tendency of the device towards a multi-mode operation with simultaneous emission in the two counter-propagating edge modes. Further islands of dynamical stability can then be found, scattered across the wide parameter space of the problem. }

\iac{In spite of these complications, some useful trends can be identified in view of ensuring a stable topolaser operation. The arguments in Sec.~\ref{sec:nonl} can be used to exploit the nonlinearity to stabilize excitation modes at small $k$ and avoid modulational instabilities. Further stabilization against the processes discussed in Sec.~\ref{sec:interplay} can be obtained with a suitable design of the lattice to further reinforce the $k$-dependence of the edge-mode localization as discussed in the earlier parts of this work, and/or of the Q-factor of the edge modes as discussed in~\cite{noh2020}.}

\iac{Guaranteeing the stability of the counter-propagating modes at $k\sim \pi$ and avoid multi-mode laser operation is instead a more subtle issue as its (typically small) imaginary part is strongly affected by the slow carrier dynamics as pointed out in Sec.~\ref{sec:slow_res} and also strongly depends on the microscopic distortion of the edge mode wavefunction by gain and nonlinearities. While these features are not easily controlled without a detailed microscopic modelling of the device, some general trends can anyway be extracted: counter-propagating mode lying in a different topological gap are well separated in frequency and therefore can be suppressed through a relatively weak frequency-dependence of gain~\cite{secli2021}.
An even more drastic strategy would be to adopt an underlying topological model that displays a single topological gap, like the Haldane model considered in~\cite{zapletal2020}.}

\section{Conclusion}
\label{sec:conclu}

\iac{In this work we have presented a general theory of the fluctuation dynamics of a topological laser device. By extending the Bogoliubov theory of the collective excitations on top of a dilute condensate to the case of a photonic topological Harper-Hofstadter model with gain and losses, we calculated the spectrum of collective excitation modes around the lasing steady-state and identified different mechanisms for dynamical instability. 
The numerical results obtained from the full 2D model were then analytically interpreted in terms of a simple effective 1D theory which only requires the knowledge of the edge mode dispersion and of their spatial localization on the edge.}

\iac{While stable topolaser operation is recovered for the class-A laser models with no optical nonlinearities considered in~\cite{amelio2019b,secli2019,zapletal2020}, more complex phenomena are found for class-B models featuring a slow carrier dynamics and/or in the presence of an intensity-dependent refractive index. Several processes potentially leading to instabilities were identified and explained in physical terms such as long-wavelength modulational instabilities or multi-mode lasing into the counter-propagating edge mode with opposite chirality. This provides physical insight into the instabilities  originally predicted with numerical tools in~\cite{longhi2018}. Based on our understanding of the various instability processes, strategies to reinforce the stability of topolaser operation are finally suggested}.


\iac{As future perspectives for further work, our study has already provided a microscopic support to the recently published theory of the long-distance, late-time coherence properties of the emission of a topological laser device~\cite{amelio2019b}. 
Concerning the recent experimental realizations of topological lasers, we agree that a quantitative study of actual semiconductor laser devices~\cite{bahari2017,bandres2018,zeng2020electrically} may require a more sophisticated modeling of the gain medium which goes beyond the scope of this work, but we anticipate that our method should be quantitatively accurate for the specific case of topological exciton-polariton condensates which are presently under active study~\cite{klembt2018}. From a conceptual perspective, we also expect that our framework will be a useful starting point for qualitative considerations and theoretical arguments.}

\begin{acknowledgments}
We acknowledge stimulating exchanges with Moti Segev and useful discussions with Stefano Longhi. We acknowledge financial support from the European Union H2020-FETFLAG-2018-2020 project "PhoQuS" (n.820392) and from the Provincia Autonoma di Trento. A. L.--P. thanks Ecole Normale Supérieure de Paris-Saclay for the stipend allocated. 
\end{acknowledgments}

\appendix*
\section{First-order corrections in $\gamma/J$}
\label{app:gamma/J}

In this Appendix, we briefly inspect the first order $\gamma/J$ corrections to the ansatz (\ref{eq:1D_ansatz},\ref{eq:ss_field_mu}-\ref{eq:ss_field_intensity}) for the lasing steady-state $\psi_y^0$ for vanishing nonlinearities $g=g_R=0$. We focus on the region in the vicinity of the edge, namely for $y=1,2$. Numerical results for the steady-state suggest that
\begin{eqnarray}
\psi^{0}_1 &\simeq & A \phi_1
\label{eq:ss_shape_0} \\
\psi^{0}_2 &\simeq & A \phi_2+ i A \frac{\gamma}{2J} \frac{1-|\phi_1|^2}{\phi_1} + O\left[\left(\frac{\gamma}{J}\right)^2\right]\, .
\label{eq:ss_shape_1}
\end{eqnarray}
While the conservative eigenstate $\{ \phi_y \}$ has purely real entries indicating the absence of particle flux in the transverse direction orthogonal to the edge, a non-trivial phase appears in the driven-dissipative steady-state due to the first order correction in $\gamma/J$. 

This interesting feature can be understood in terms of particle conservation at the different sites.
The total radiative losses of the lasing mode (per unit length along $x$) are given by 
\begin{equation}
\gamma \sum_y |\psi^{0}_y|^2\simeq \gamma A^2
\end{equation}
where the second equality holds up to second order corrections. All gain is concentrated on the $y=1$ sites, while the contribution of this sites to the losses
\begin{equation}
    \gamma |\psi^{0}_1| =  \gamma A^2 |\phi^{0}_1|^2
\end{equation} 
is only partial.
Overall balance between gain and losses then requires the presence of some current flowing out of the edge, namely from $y=1$ to $y=2$. This current is provided exactly by the first order correction included in Eq.~(\ref{eq:ss_shape_1}): even if this term is order $\gamma/J$ the associated current contains the tunnelling rate $J$ and is of order $\gamma$. 
Including this current, the total rate of particle escape from the $y=1$ sites recovers that of the whole lattice $\gamma A^2 |\phi_1|^2$. 

\newcounter{preveqcounter}
\setcounter{preveqcounter}{\value{equation}}

\section{Two-dimensional Bogoliubov theory including the carrier dynamics}\label{app:Bogo2D}

\setcounter{equation}{\value{preveqcounter}}

Including carrier dynamics, the Ansatz (\ref{eq:dpsi}) becomes:

\begin{equation}\label{eq:dpsi_app}
\left\{ 
    \begin{array}{lll}
    \psi_{x,y}(t) = [\psi^0_y +  \delta \psi_{x,y}(t)]\, e^{-i\omega^{\textrm{Las}} t + i k_{x}^{\textrm{Las}}x}\\
    \\
    N_{x,y}(t) = N_{y}^0 + \delta  N_{x,y}(t)
    \end{array}
\right.
\end{equation}

In Fourier space along the $x-$axis and time, the corresponding system of linear equations for 
\iac{$\delta\psi_{k,y}$ and $\delta  N_{k,y}$ reads for $k=k_x-k_{x}^{\textrm{Las}}$:}
\begin{equation}\label{eq:HH2D_ad_lin_app}
\left\{ 
    \begin{array}{lll}
    \begin{aligned}    
      \omega^{\rm Bog}(k)&\,\delta \psi_{k,y} = (\mathbf{H}\,\delta \psi)_{k,y}  + g\psi^0_y \ ^2 \delta \psi^*_{y,-k} \\
    &+[2g|\psi^0_y|^2 + g_R N_y^0 - \omega^{\textrm{Las}} + i(R N^0_y - \gamma)  ]\delta \psi_{y,k} \\
    & + \psi^0_y[i R + g_R]\delta N_{y,k}
    \end{aligned} \\
    \\
    \begin{aligned}
    \omega^{\rm Bog}(k)\,\delta N_{k,y} =& - i(\gamma_R + R|\psi^0_{y}|^2)\delta N_{k,y}\\
    & - R N^0_y ( \psi^0_y \delta \psi^*_{-k,y} + \psi^{0*}_y \delta \psi_{k,y})  
    \end{aligned}
    \end{array}
\right. ,    
\end{equation}
to be supplemented for 
\iac{$\delta\psi_{k,y}^*$}
by the complex conjugate of the first equation.

\bibliographystyle{apsrev4-1}   
\bibliography{bibliography}

\end{document}